\documentclass[preprint,superscriptaddress,amsmath,amssymb,aps,prf,longbibliography]{revtex4-1}

\usepackage{graphicx}
\usepackage{epstopdf, epsfig}
\usepackage{dcolumn}
\usepackage{bm}
\usepackage{gensymb}
\usepackage{enumitem}
\usepackage{mathrsfs}
\usepackage[sort&compress]{natbib}
\usepackage{color}
\usepackage{hyperref}

\begin{document}


\title{On skin friction in wall-bounded turbulence}

\author{Zhenhua Xia}
\email{xiazh@zju.edu.cn}
\affiliation{Department of Engineering Mechanics, Zhejiang University, Hangzhou 310027, China}
\author{Peng Zhang}
\affiliation{Department of Engineering Mechanics, Zhejiang University, Hangzhou 310027, China}
\author{Xiang I. A. Yang}
\affiliation{Mechanical Engineering, Pennsylvania State University, State College, Pennsylvania 16802, USA}

\date{\today}
\begin{abstract}
In this paper, we derive mathematical formulas for the skin friction coefficient in wall-bounded turbulence based on the Reynolds averaged streamwise momentum equation and the total stress.
Specially, with the theoretical or empirical relation of the total stress, the skin friction coefficient is expressed in terms of the mean velocity and the Reynolds shear stress in an arbitrary wall-normal region $[h_0, h_1]$.
The formulas are validated using the direct numerical simulation data of turbulent channel and boundary layer flows, and the results show that our formulas estimate the skin friction coefficient very accurately with an error less than $2\%$. We believe that the present integral formula can be used to determine the skin friction in turbulent channel and boundary layer flows at high Reynolds numbers where the near wall statistics are very difficult to measure accurately.
\end{abstract}

\maketitle

\section{Introduction}

Wall-bounded turbulence is ubiquitous in nature and engineering applications.
In these flows, the skin friction $\tau_w$, or the wall shear stress, is of great importance to both practical engineering and fundamental fluid physics.
From engineering view, it was estimated that the skin friction drag might account for around $50\%$ and $30\%$ of the total drag for a long-range subsonic airplane~\cite{Kornilov2005} and a ground vehicle~\cite{Patten2012} respectively. From the view of fluid physics, knowing the wall shear stress is vital to determine the law of the velocity profiles in wall-bounded turbulence. Therefore, determining the skin friction accurately is an important issue for wall-bounded turbulence, and it is also a challenge for the community~\cite{Naughton2002}. For examples, if the wall shear stress is determined through the mean velocity gradient near the wall experimentally, then the accuracy will be limited by the spatial resolution of the measurement system. On the other hand, if it is estimated by using the electro-chemical methods or the oil-film interferometry (OFI), then the accuracy will depend on the wall condition and fluid type~\cite{Winter1977, Tavoularis2005, Mehdi2011, Zanoun2014, Segalini2015}. Other experimental methods, such as the hot-wire anemometry (HWA)~\cite{Hutchins2002}, the laser-Doppler anemometry (LDA)~\cite{Zanoun2014}, can also be used to estimate the skin friction. However, they also suffers from some limitations~\cite{Ikeya2017,Niegodajew2019}.

An alternative to those experimental methods mentioned above is the Clauser-chart method~\cite{Clauser1954,Zanoun2014,Niegodajew2019}. In this method, the universal logarithmic behavior of the mean velocity profile,
\begin{equation}\label{eq:loglaw}
\frac{U}{u_\tau}=\frac{1}{\kappa}\ln(\frac{yu_\tau}{\nu})+C,
\end{equation}
is assumed and adopted with $\kappa$ being the von Karman constant and $C$ being the additive constant. Here, $u_\tau=\sqrt{\tau_w/\rho}$ is the friction velocity at the wall with $\rho$ being the fluid density, $\nu$ is the kinematic viscosity and $U$ is the mean velocity at the a wall-normal displacement $y$. With known $\kappa$ and $C$, the friction velocity $u_\tau$ can be obtained by fitting the measured mean velocity to equation~\eqref{eq:loglaw}, and then the skin friction can be obtained according to $\tau_w=\rho u_\tau^2$. Nevertheless, the proper estimation of $u_\tau$ will largely depend on the values of $\kappa$ and $C$, and it was concluded by Crook~\cite{Crook2002} that a $\pm 0.5$ change in the slope $1/\kappa$ will result in a $12\%$ difference in $u_\tau$. Furthermore, it is clearly that severe error may exist if the Clauser-chart method is used in the cases where the logarithmic behavior of the mean velocity is invalid. In this situation, the classic Clauser-chart method has to be modified or corrected~\cite{Niegodajew2019}.

There are other methods to determine the skin friction, where integral form is adopted. Ligrani and Moffat~\cite{Ligrani1986} used the momentum integration equation to determine the skin friction. However, this method involves the integration of the streamwise gradient terms, which are often difficult to get from experimental data, and it is unusable if the profiles at multiple streamwise locations are not available. Mehdi and White~\cite{Mehdi2011} proposed an integration form of the skin friction coefficient based on the FIK decomposition~\cite{FIK2002} which is suitable for experimental data. Differently from the original FIK decomposition, where a streamwise inhomogeneous term exists for the boundary layer flow and it also involves the streamwise gradient terms, Mehdi and White~\cite{Mehdi2011} introduced the total stress gradient term to replace the streamwise inhomogeneous term and the pressure gradient term, which are mathematically equivalent. The total stress can be determined based on the profiles of the mean velocity gradient and the Reynolds shear stress, and then the new integral form can be used to estimate the skin friction with experimental data acquired at only one streamwise location and within at least one $\delta$. Here, $\delta$ is the boundary layer thickness pertaining to $0.99$ free stream velocity. Later on, Mehdi {\it et al.}~\cite{Mehdi2014} found that the FIK decomposition was based on an exact equation and thus the integration could be performed to an arbitrary height. The resulted integral formula can be used to flows with ill-defined outer boundary conditions or when the measurement grid does not extend over the whole boundary layer thickness.

Nevertheless, the above integral formula of the skin friction from Mehdi {\it et al.}~\cite{Mehdi2014} contains $(1-y)$ and $(1-y)^2$ weightings for the Reynolds shear stress term and the total stress gradient term respectively, and these weightings place more emphasis on the near-wall values of the Reynolds shear stress and the total stress gradient. Based on the direct numerical simulation data at $Re_\theta=4060$ from Schlatter and \"Orl\"{u}~\cite{Schlatter2010}, Mehdi {\it et al.}~\cite{Mehdi2014} showed that neglecting the statistics within $y/\delta<0.01$ will result in more than $2\%$ error in the skin friction when the upper boundary of the integration is around $0.3\delta$ and this error will increase to $8\%$ if the lower boundary of the integration increases to $y/\delta=0.02$. This fact demonstrates the importance of the near-wall statistics for the integral formula from Mehdi {\it et al.}. On the other hand, the near-wall statistics will largely limited by the spatial resolution of the measurement system and an accurate measurement of the statistics is also a challenge by itself~\cite{Zhu2013}. Therefore, an estimation method which is based sorely on the turbulence statistics away from the wall at one single streamwise location is in need, and this is the main focus of the present work.

The remainder of the paper is organized as follows. Section~\ref{sec:Math} will present the mathematical formulation of our new method and its detailed form for incompressible turbulent channel flows and boundary layers. The validations for incompressible turbulent channel flows and boundary layers will be presented in Section~\ref{sec:results}, followed by conclusion in Section~\ref{sec:Con}.

\section{Mathematical formulation} \label{sec:Math}

The starting point of the formula is the Reynolds averaged $x-$momentum equation (i.e., the equation in the streamwise direction). For a statistically stationary two-dimensional wall-bounded flow, the Reynolds averaged $x-$momentum equation is given by
\begin{equation}\label{eq:x-mom}
0=\frac{\partial}{\partial y}\left[\overline{u'v'}-\frac{1}{Re}\frac{\partial \bar{u}}{\partial y}\right] + \overline{I_x} + \frac{d \bar{P}}{d x},
\end{equation}
where the equation is normalized by the free stream velocity $U_\infty$ and the boundary thickness $\delta$. Here, $y$ is the wall-normal direction, $u, v$ are the velocity components in the streamwise and wall-normal directions with bars denoting the mean quantities and primes denoting the corresponding fluctuation. $Re=U_\infty \delta/\nu$ is the Reynolds number based on the boundary thickness and the free stream velocity, $P$ is the pressure with the constant density absorbed and
$$\overline{I_x}=\bar{u}\frac{\partial \bar{u}}{\partial x} + \bar{v}\frac{\partial \bar{u}}{\partial y}-\frac{1}{Re}\frac{\partial^2 \bar{u}}{\partial x^2}+\frac{\partial \overline{u'^2}}{\partial x}.$$
In the following, the constant density $\rho$ is assumed to be one and it will be omitted for brevity.

As pointed out by Mehdi {\it et al.}~\cite{Mehdi2014}, the above equation~\eqref{eq:x-mom} is exact, and it is true even in separated regions. Following the procedure by Fukagata {\it et al.}~\cite{FIK2002}, integrating equation~\eqref{eq:x-mom} three times in the wall-normal direction, one may obtain the mathematical relationship for the skin friction coefficient, $C_f=\tau_w/(\frac{1}{2} U_\infty^2)$,
which reads~\cite{Mehdi2011},
\begin{equation}\label{eq:FIK}
C_f=\frac{4(1-\delta^*)}{Re} +2\int_0^12(1-y)(-\overline{u'v'})dy+2\int_0^1(1-y)^2(-\overline{I_x})dy-\frac{2}{3}\frac{d \bar{P}}{d x},
\end{equation}
where $\delta^*$ is the displacement thickness normalized using $\delta$. By introducing the total stress
\begin{equation}\label{eq:tau}
\tau=\frac{1}{Re}\frac{\partial \bar{u}}{\partial y}-\overline{u'v'},
\end{equation}
Mehdi and White~\cite{Mehdi2011} obtained the following expression for the skin friction coefficient
\begin{equation}\label{eq:Mehdi1}
C_f=\frac{4(1-\delta^*)}{Re}+2\int_0^1 2(1-y)(-\overline{u'v'})dy+2\int_0^1(1-y)^2\left(-\frac{\partial \tau}{\partial y}\right)dy.
\end{equation}
In their paper~\cite{Mehdi2011}, Mehdi and White fit a Whittaker smoother with a small smoothing parameter to the data of $(1-y)\tau$ ($0 \le y \le 1$) with the condition that the gradient of the fitting function $g(y)$ always remains negative. By doing so, they can estimate $\tau=g(y)/(1-y)$ and then obtain its gradient. Later on, Mehdi {\it et al.}~\cite{Mehdi2014} proposed to integrate equation~\eqref{eq:x-mom} to an arbitrary height $y_t$ three times and with the help of the total stress, they obtained the expression for the skin friction coefficient using the information for $y<y_t$,
\begin{equation}\label{eq:Mehdi2}
C_f=\frac{4}{y_t^2} \left[\frac{1}{Re}\int_0^{y_t}\bar{u}dy -\int_0^{y_t}(y_t-y)\overline{u'v'}dy-\frac{1}{2}\int_0^{y_t}(y_t-y)^2\frac{\partial \tau}{\partial y}dy\right].
\end{equation}

In fact, if the behavior of $\tau$ as a function of $y$ is known, then the skin friction can be obtained simply by setting $y=0$. Alternatively, we could integrate equation~\eqref{eq:x-mom} to an arbitrary height $y_t$ twice, and we can obtain another expression for the skin friction coefficient
\begin{equation}\label{eq:XZH1}
C_f=\frac{2}{y_t} \left[\frac{1}{Re}\bar{u}(y_t) -\int_0^{y_t}\overline{u'v'}dy-\int_0^{y_t}(y_t-y)\frac{\partial \tau}{\partial y}dy\right],
\end{equation}
where the non-slip boundary condition on the streamwise velocity component is applied, i.e., $\bar{u}(0)=0$. Clearly, equation~\eqref{eq:XZH1} can also be used to determine the skin friction coefficient, and its dependence on the near-wall statistics is weaker than equation~\eqref{eq:Mehdi2}. These two equations can both be used to estimate the skin friction coefficient for general wall-bounded turbulent flows.

\subsection{Formula for turbulent channel flow}
For an statistically stationary incompressible turbulent channel flow, it is well known that the total stress $\tau$ decays linearly with the wall distance as
\begin{equation}\label{eq:Chan-tau}
\tau=\tau_w(1-y).
\end{equation}
Or
\begin{equation}\label{eq:Chan-tau2}
\frac{1}{8}(1-y)C_f=\frac{1}{Re_b}\frac{\partial \bar{u}}{\partial y}-\overline{u'v'},
\end{equation}
where all the quantities are normalized by $2U_b$ and $h$ with $U_b$ and $h$ being the bulk velocity and half of the channel width respectively, and $Re_b=2U_bh/\nu$ is the bulk Reynolds number. Here, equation~\eqref{eq:Chan-tau2} is exact and it is true for any $y$. From equation~\eqref{eq:Chan-tau2},
we could obtain an expression for the skin friction coefficient as
\begin{equation}\label{eq:Chan-1}
C_f=\left.\frac{8}{1-y}\right[\left.\frac{1}{Re_b}\frac{\partial \bar{u}}{\partial y}-\overline{u'v'}\right].
\end{equation}
Hence, $C_f$ could be evaluated from equation~\eqref{eq:Chan-1} at any arbitrary wall-normal location $y$ except the channel center where $y=1$.

Furthermore, when equation \eqref{eq:Chan-tau} is inserted into equation~\eqref{eq:Mehdi2} and~\eqref{eq:XZH1}, simplified expressions containing only the mean velocity and Reynolds shear stress, can be obtained, where the integration starts from the wall to an arbitrary height. Nevertheless, as equation~\eqref{eq:Chan-tau2} is exact at any $y$, we can integrate it from an arbitrary start $h_0$ to another arbitrary end $h_1$ once or twice and then we will arrive at the following expressions
\begin{equation}\label{eq:Chan-2}
C_f=\left.\frac{1}{A_1}\right[\left.\frac{1}{Re_b}(\bar{u}(h_1)-\bar{u}(h_0))-\int_{h_0}^{h_1}\overline{u'v'}dy\right],
\end{equation}
 and
\begin{equation}\label{eq:Chan-3}
C_f=\left.\frac{1}{A_2}\right[\left.\frac{1}{Re_b}\int_{h_0}^{h_1}\bar{u}dy-\frac{\bar{u}(h_0)}{Re_b}(h_1-h_0)-\int_{h_0}^{h_1}(h_1-y)\overline{u'v'}dy\right].
\end{equation}
Here,
\begin{equation*}
A_1=\frac{(2-h_1-h_0)(h_1-h_0)}{16},\quad A_2=\frac{(3-h_1-2h_0)(h_1-h_0)^2}{48}.
\end{equation*}
In equation~\eqref{eq:Chan-3}, if we set $h_0=0$ and $h_1=1$, then it will recover the FIK decomposition~\cite{FIK2002}. More importantly, equation~\eqref{eq:Chan-2} and~\eqref{eq:Chan-3} do not require the information from the near-wall and it can be used to estimate the skin friction coefficient when the near-wall statistics is absent.

\subsection{Formula for turbulent boundary layer}

For turbulent boundary layers, there is not an analytical expression for the total stress. Fortunately, Hou {\it et al.}~\cite{Hou2006} found that a very simple and accurate linear fit exists for $y<0.5$ in zero pressure gradient turbulent boundary layers when the total stress profile is weighted with $(1-y)$, i.e.,
\begin{equation}\label{eq:BL-linear}
    (1-y)\frac{\tau}{\tau_w}=ay+b \quad for\quad y\le 0.5.
\end{equation}
In their paper~\cite{Hou2006}, they found that the value of $a$ is within the range $-1.6\sim-1.2$ and there is no clear dependence on the Reynolds number. Theoretically, $b=1$ since $\tau=\tau_w$ at the wall. The value of $a$ could be determined through the data. 
With this empirical relationship, we could obtain some similar formulas for the turbulent boundary layer flow with zero pressure gradient. From equation~\eqref{eq:BL-linear} and the definition of total stress, we could obtain the following relation for the skin friction coefficient
\begin{equation}\label{eq:BL-1}
   C_f=\left.\frac{2(1-y)}{ay+1}\right[\left.\frac{1}{Re}\frac{\partial \bar{u}}{\partial y}-\overline{u'v'}\right].
\end{equation}
Alternatively, we could integrate equation~\eqref{eq:BL-1} from an arbitrary start $h_0$ to another arbitrary end $h_1$ once or twice, and then we will get the following expressions for the skin friction coefficient
\begin{equation}\label{eq:BL-2}
C_f=\left.\frac{1}{B_1}\right[\left.\frac{1}{Re}(\bar{u}(h_1)-\bar{u}(h_0))-\int_{h_0}^{h_1}\overline{u'v'}dy\right],
\end{equation}
 and
\begin{equation}\label{eq:BL-3}
C_f=\left.\frac{1}{B_2}\right[\left.\frac{1}{Re}\int_{h_0}^{h_1}\bar{u}dy-\frac{\bar{u}(h_0)}{Re}(h_1-h_0)-\int_{h_0}^{h_1}(h_1-y)\overline{u'v'}dy\right].
\end{equation}
Here,
$$B_1=\frac{-a(h_1-h_0)+(a+1)(\ln(1-h_0)-\ln(1-h_1))}{2},$$ 
$$B_2=\frac{-a(h_1-h_0)^2+2(a+1)(h_1-h_0)+2(a+1)(h_1-1)(\ln(1-h_0)-\ln(1-h_1))}{4}.$$
Note that in equations~\eqref{eq:BL-2} and~\eqref{eq:BL-3}, $0\le h_0 < h_1\le 0.5$, while in equation~\eqref{eq:BL-1}, $0\le y\le 0.5$ and $y \ne -1/a$. Furthermore, equations~\eqref{eq:BL-2} and~\eqref{eq:BL-3} are exactly equivalent to equation~\eqref{eq:XZH1} and~\eqref{eq:Mehdi2} respectively when the total stress form~\eqref{eq:BL-linear} is used with $h_0=0$ and $h_1=y_t\le 0.5$.

\section{Results} \label{sec:results}
\subsection{Validation in turbulent channel flow}

\begin{table}
  \begin{center}
  \setlength{\tabcolsep}{3.5mm}
  \begin{tabular}{ccccccccc}
  \hline
  case	  & $Re_\tau$ & $Re_b$ & $L_x/h$ & $L_z/h$ & $N_x$	& $N_y$ & $N_z$ &$C_f^0$    \\
  \hline
  CH180  & 182.1 & 5714.3  & $8\pi$ & $3\pi$	 & 1024 & 192 & 512 & $8.123\times 10^{-3}$ \\
  CH550  & 543.5 & 20000  & $8\pi$ & $3\pi$	 & 1536 & 384 & 1024 & $5.908\times 10^{-3}$ \\
  CH1000  & 1000.5 & 40000  & $8\pi$ & $3\pi$	 & 2304 & 512 & 2048 & $5.005\times 10^{-3}$ \\
  CH2000  & 1994.8 & 86956  & $8\pi$ & $3\pi$	 & 4096 & 768 & 3072 & $4.210\times 10^{-3}$ \\
  CH5200  & 5185.9 & 250000  & $8\pi$ & $3\pi$	 & 10240 & 1536 & 7680 & $3.442\times 10^{-3}$ \\
  \hline
  \end{tabular}
  \caption{Parameters about the direct numerical simulations from Lee and Moser~\cite{Lee2015}. Here, the parameters are gathered from the data files downloaded from http://turbulence.ices.utexas.edu. Reynolds numbers are defined as $Re_\tau=u_\tau h/\nu$, $Re_b=2U_bh/\nu=Re_\tau*(2U_b/u_\tau)$. The reference skin friction is calculated through $C_f^0=\tau_w/(\rho U_b^2/2)=2(u_\tau/U_b)^2=8(Re_\tau/Re_b)^2$.}
  \label{tab:1}
  \end{center}
\end{table}

In this subsection, we validate the formulas of the skin friction coefficient in turbulent channel flow. The direct numerical simulation data at several different Reynolds numbers from Lee and Moser~\cite{Lee2015} will be used. The detailed parameters are listed in Table~\ref{tab:1}. Here, we mainly validate equations~\eqref{eq:Chan-1},~\eqref{eq:Chan-2} and \eqref{eq:Chan-3} instead of equations~\eqref{eq:XZH1} and~\eqref{eq:Mehdi2}, since the latter two equations can be recovered through equations~\eqref{eq:Chan-2} and \eqref{eq:Chan-3} by setting $h_0=0$. Trapezoidal rule is adopted to estimate the integrals in equations~\eqref{eq:Chan-2} and \eqref{eq:Chan-3}, while the gradient of the mean velocity is already included in the data-sets.

\begin{figure}
    \centering
    \includegraphics[width=0.85\textwidth]{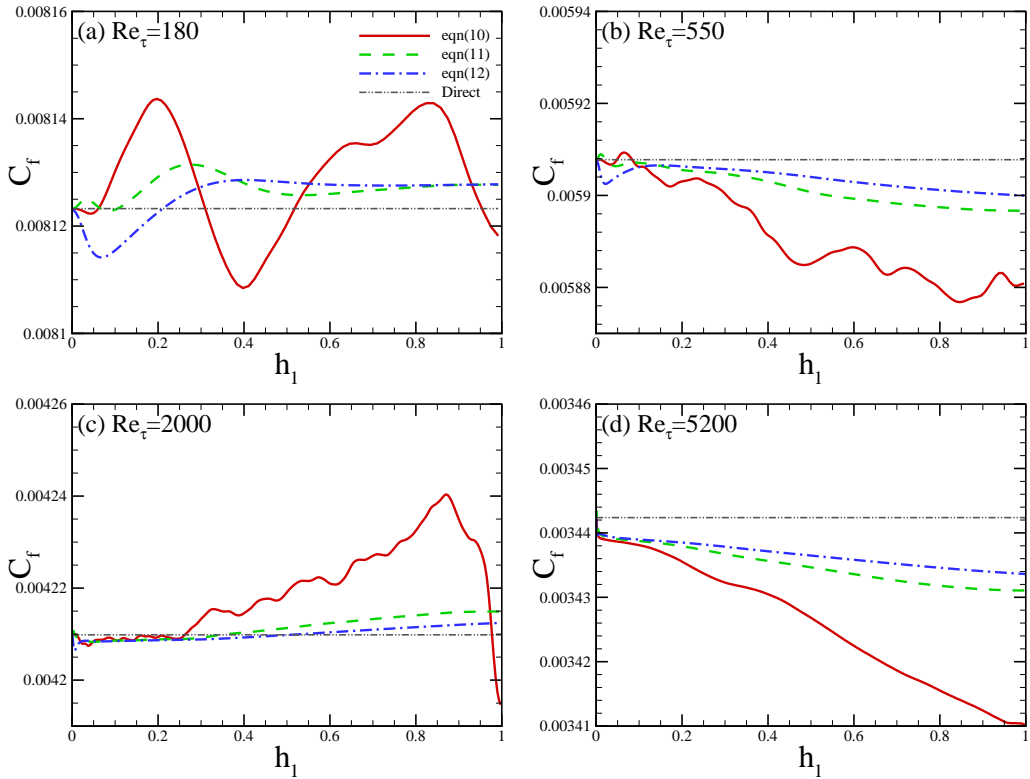}
    \caption{Skin friction coefficient estimated using equations~\eqref{eq:Chan-1},~\eqref{eq:Chan-2} and \eqref{eq:Chan-3} respectively. For equation~\eqref{eq:Chan-1}, $y=h_1$. For equations~\eqref{eq:Chan-2} and \eqref{eq:Chan-3}, $h_0=0$. (a) CH180; (b) CH550; (c) CH2000 and (d) CH5200. The horizontal dashed-double-dotted line is the corresponding reference $C_f^0$.}
    \label{fig:CH-Cf}
\end{figure}

\begin{figure}
    \centering
    \includegraphics[width=0.96\textwidth]{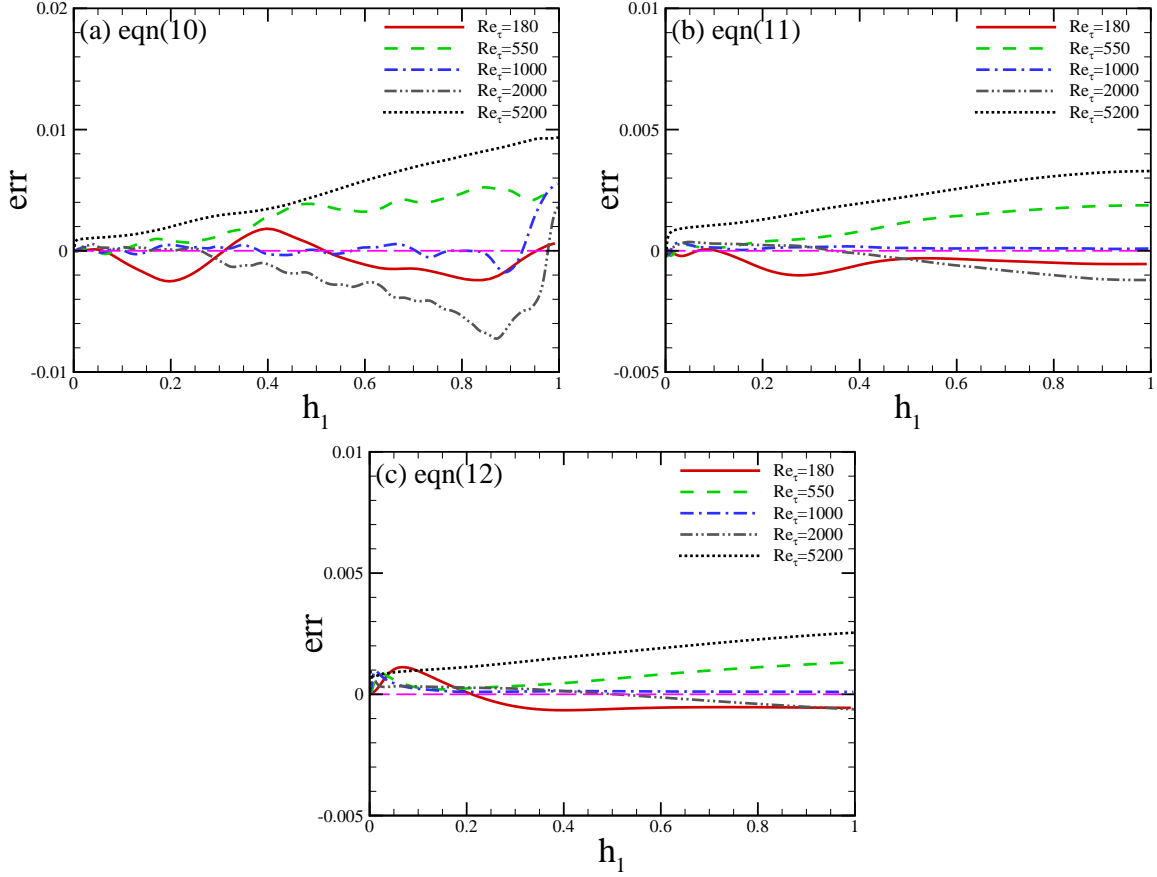}
    \caption{Relative error of the skin friction coefficient at different Reynolds numbers with $h_0=0$. (a) Using equation~\eqref{eq:Chan-1}; (b) Using equation~\eqref{eq:Chan-2} and (c) Using equation~\eqref{eq:Chan-3}.}
    \label{fig:CH-Cf-err}
\end{figure}

\begin{figure}
    \centering
    \includegraphics[width=0.47\textwidth]{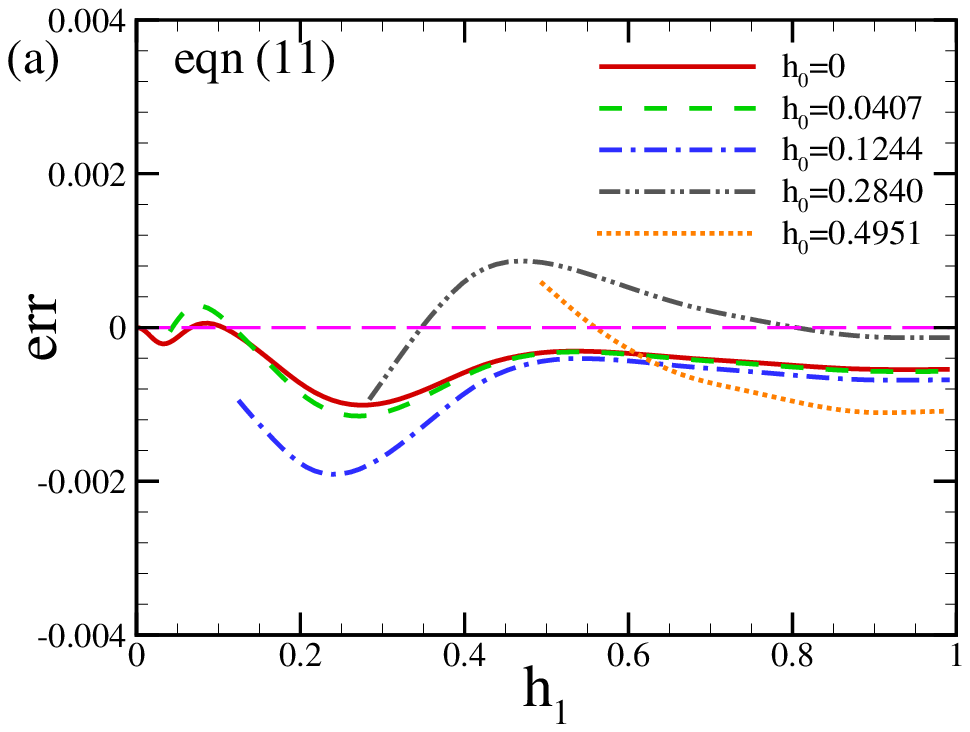}\includegraphics[width=0.47\textwidth]{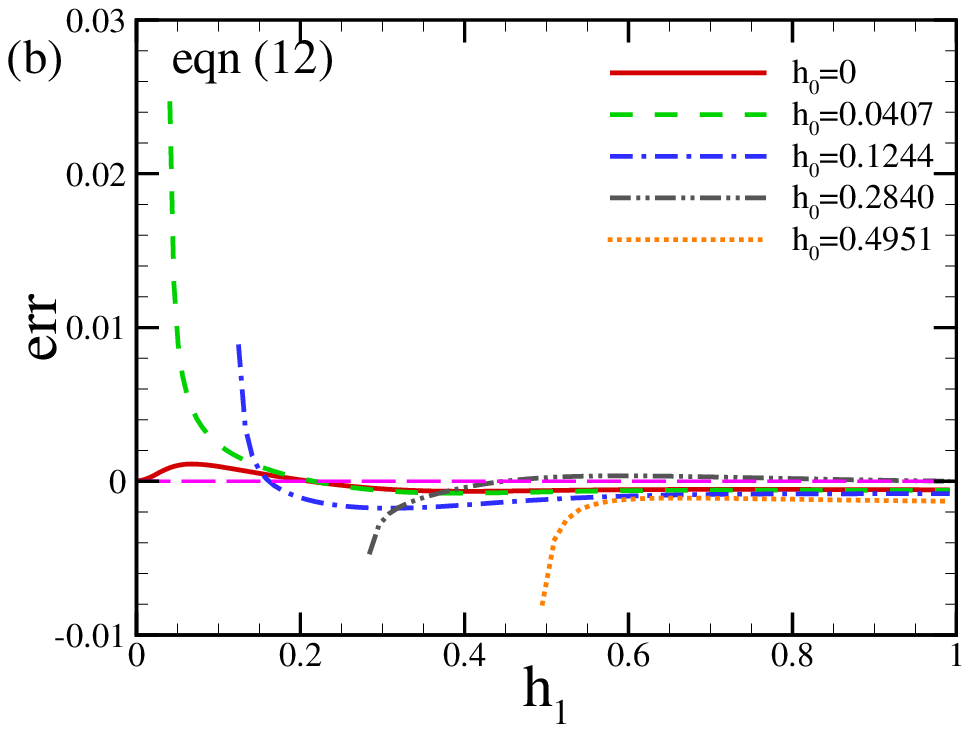}\\
   \includegraphics[width=0.47\textwidth]{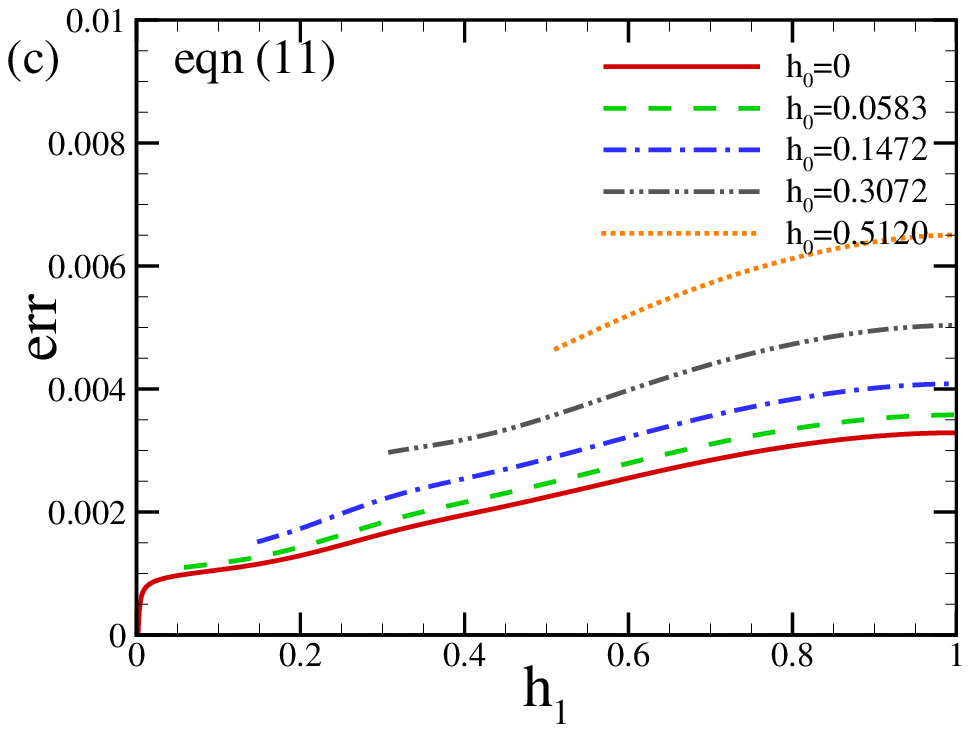}\includegraphics[width=0.47\textwidth]{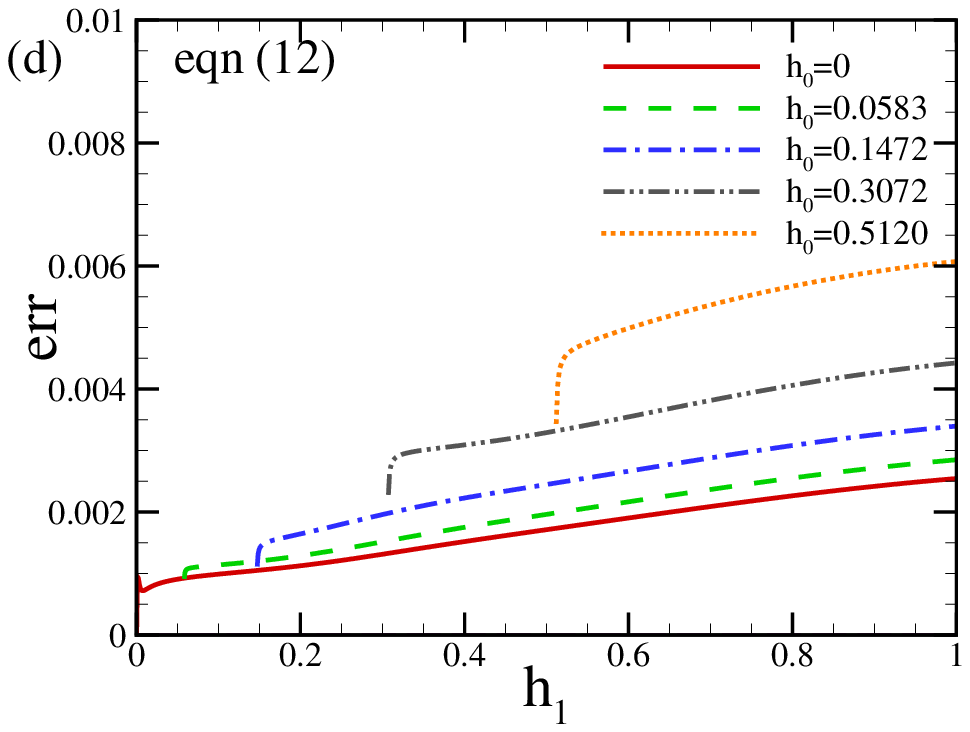}
    \caption{ Related error of the skin friction coefficient estimated using equations~\eqref{eq:Chan-2} and~\eqref{eq:Chan-3} with different $h_0$ for case CH180 (a,b) and CH5200 (c,d). (a) and (c) are using equation~\eqref{eq:Chan-2}; (b) and (d) are using equation~\eqref{eq:Chan-3}. }
    \label{fig:CH-180}
\end{figure}

Figure~\ref{fig:CH-Cf} shows the skin friction coefficient estimated using equations~\eqref{eq:Chan-1},~\eqref{eq:Chan-2} and \eqref{eq:Chan-3} at four different Reynolds numbers. For equations~\eqref{eq:Chan-2} and \eqref{eq:Chan-3}, the integration starts at the wall with $h_0=0$. It is seen that the estimations at any fixed Reynolds numbers from all of the three methods are very close to the corresponding reference value and the deviations from equations~\eqref{eq:Chan-2} and \eqref{eq:Chan-3} are generally smaller than those from equation\eqref{eq:Chan-1}. This is consistent with the fact that equation \eqref{eq:Chan-1} is mainly based on the balance of the mean streamwise momentum and it is locally dependent on the mean velocity gradient and the Reynolds shear stress. On the other hand, equations~\eqref{eq:Chan-2} and \eqref{eq:Chan-3} are integrals of equation \eqref{eq:Chan-1}. For equation~\eqref{eq:Chan-2}, the estimation depends on the local mean velocity differences between two ends and the average of the Reynolds shear stress in the integrating range. For equation~\eqref{eq:Chan-3}, the estimation depends on the local mean velocity at the integrating start point, the local mean velocity (or mass flux) at the integrating range and the weighted average of the Reynolds shear stress in the integrating range. It can be observed from figure~\ref{fig:CH-Cf} that the deviations from the reference values by using equations~\eqref{eq:Chan-2} and \eqref{eq:Chan-3} do not become smaller as $h_1$ increases to $1.0$. This could be explained by the slight unbalance of the mean streamwise momentum, which are also shown in the figure. Nevertheless, the present data are very accurate, and thus the errors using the three different formulas are very small, as shown in Figure~\ref{fig:CH-Cf-err}, where the related error of the estimations using equations~\eqref{eq:Chan-1}, \eqref{eq:Chan-2} and \eqref{eq:Chan-3} with $h_0=0$ at five different Reynolds numbers are shown respectively. Here, the error is defined as
\begin{equation}
err = \frac{C_f^0-C_f}{C_f^0}=1- \frac{C_f}{C_f^0}.
\end{equation}
It is seen from figure~\ref{fig:CH-Cf-err} that the errors are all very small, whereas the errors using equation~\eqref{eq:Chan-1} are within $1\%$, while the errors using equations~\eqref{eq:Chan-2} and \eqref{eq:Chan-3} are within $0.5\%$.

Now, we are going to investigate the estimations using equations~\eqref{eq:Chan-2} and \eqref{eq:Chan-3} with different $h_0$ and $h_1$. In figure~\ref{fig:CH-180}, the relative error of the skin friction coefficient estimated using equations~\eqref{eq:Chan-2} and \eqref{eq:Chan-3} with five different $h_0$ and varying $h_1$ from cases CH180 and CH5200 are shown. For case CH180, it is seem from figure~\ref{fig:CH-180} that the estimations using equations~\eqref{eq:Chan-2} and \eqref{eq:Chan-3} are very accurate for all five different values of $h_0$, verifying the correctness of equations~\eqref{eq:Chan-2} and \eqref{eq:Chan-3}. For equation~\eqref{eq:Chan-2}, the estimated skin friction coefficient at a certain location $h_1$ varies with different $h_0$. Nevertheless, the differences are quite small. The relative errors estimated using equation~\eqref{eq:Chan-2} with different $h_0$ and $h_1$ are within $0.2\%$, as depicted in figure~\ref{fig:CH-180}(a). Furthermore, the relative errors do not show any dependence on the value of $(h_1-h_0)$. Figure~\ref{fig:CH-180}(b) shows that the estimated skin friction coefficient using equation~\eqref{eq:Chan-3} can also be very accurate with different $h_0$ on the condition that the value of $(h_1-h_0)$ is large enough. At the present case CH180, it is seen that $(h_1-h_0) > 0.1$ can make sure of a relative error within $0.2\%$. At higher Reynolds number case CH5200, the estimated skin friction coefficient using equations~\eqref{eq:Chan-2} and \eqref{eq:Chan-3} are again very accurate as compared to the reference value and the relative error are all within $0.7\%$, as shown in figure~\ref{fig:CH-180}(c) and (d). Again, there is no obvious dependence on the value of $(h_1-h_0)$ for equation~\eqref{eq:Chan-2}, whereas some dependence on the value of $(h_1-h_0)$ can be observed for equation~\eqref{eq:Chan-3} when $(h_1-h_0)< 0.05$. Nevertheless, due to the imbalance of streamwise momentum at case CH5200, as shown in figure~\ref{fig:CH-Cf}(d), equations~\eqref{eq:Chan-2} and \eqref{eq:Chan-3} will both under estimate the skin friction coefficient.

\subsection{Validation in turbulent boundary layer}

In this subsection, we will validate the formulas of the skin friction coefficient in turbulent boundary layer flow. The direct numerical simulation data of a turbulent boundary layer from Schlatter and \"{O}rl\"{u}~\cite{Schlatter2010} will be used. Ten profiles at different Reynolds numbers can be downloaded from https://www.mech.kth.se/$\sim$pschlatt/DATA/README.html. For more information about the simulation, validation about the data, please refer to the reference~\cite{Schlatter2010}. Similarly, we mainly validate equations~\eqref{eq:BL-1},~\eqref{eq:BL-2} and \eqref{eq:BL-3} instead of equations~\eqref{eq:XZH1} and~\eqref{eq:Mehdi2} here. Again, the trapezoidal rule is adopted to estimate the integrals in equations~\eqref{eq:Chan-2} and \eqref{eq:Chan-3}, while the gradient of the mean velocity is already included in the data-sets.

\begin{figure}
    \centering
    \includegraphics[width=0.94\textwidth]{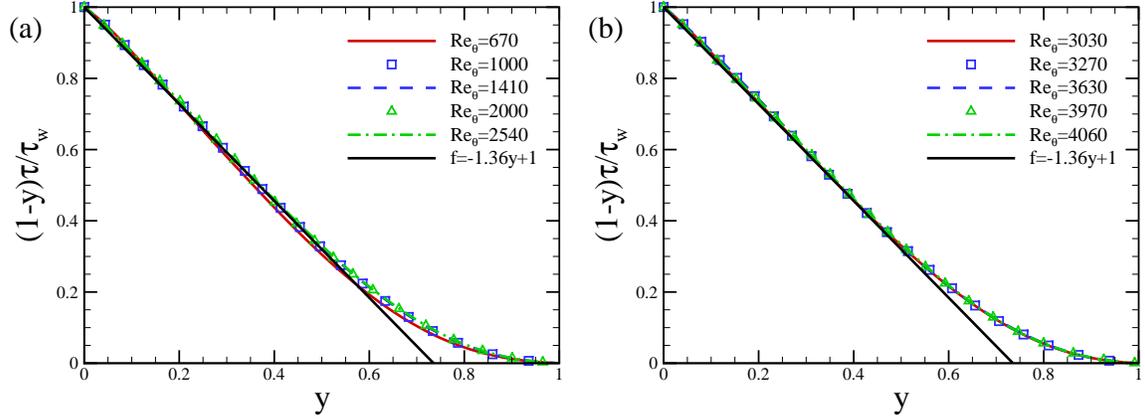}
    \caption{The profiles of $(1-y)\tau/\tau_w$ at different $Re_\theta$. (a) $670 \le Re_\theta \le 2540$; (b)  $3030 \le Re_\theta \le 4060$. A reference line $f=-1.36y+1$ is also included.}
    \label{fig:BL-tau1my}
\end{figure}

Figure~\ref{fig:BL-tau1my} shows the profiles of $(1-y)\tau/\tau_w$ at ten different streamwise locations. It is seen that $(1-y)\tau/\tau_w$ indeed follows a linear relation $f=-1.36y + 1$ for $0\leq y \leq 0.5$ at all of the ten locations, as proposed by Hou {\it et al.}~\cite{Hou2006}. In the following, we will use $a=-1.36$ in equations~\ref{eq:BL-1},~\ref{eq:BL-2} and~\ref{eq:BL-3} unless otherwise stated.

\begin{figure}
    \centering
    \includegraphics[width=0.47\textwidth]{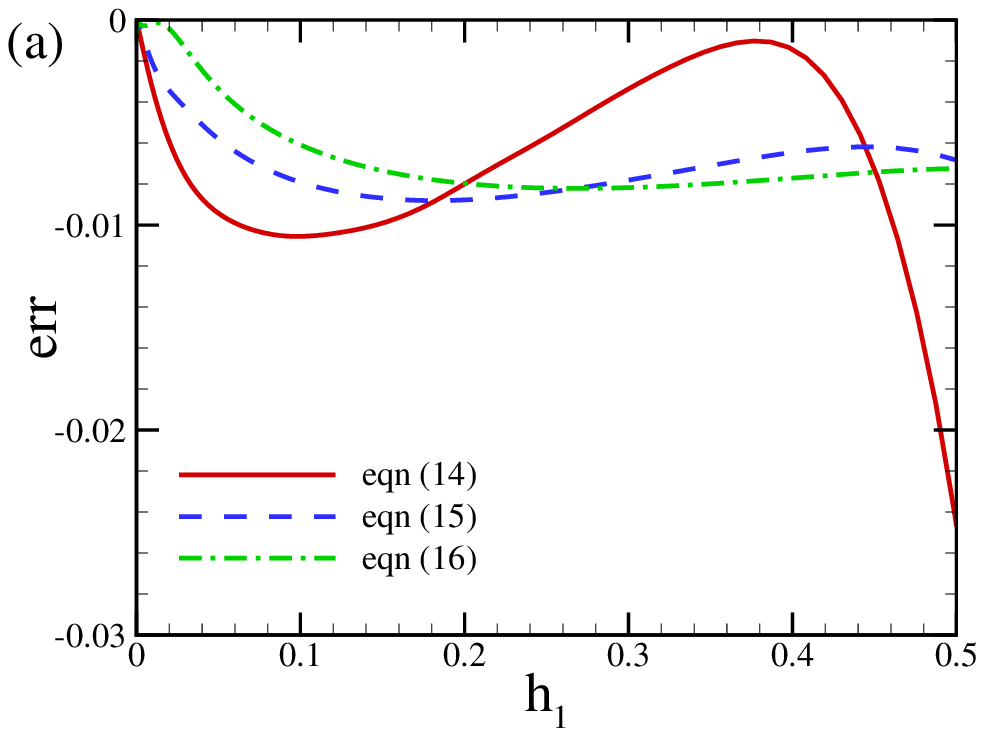}\includegraphics[width=0.47\textwidth]{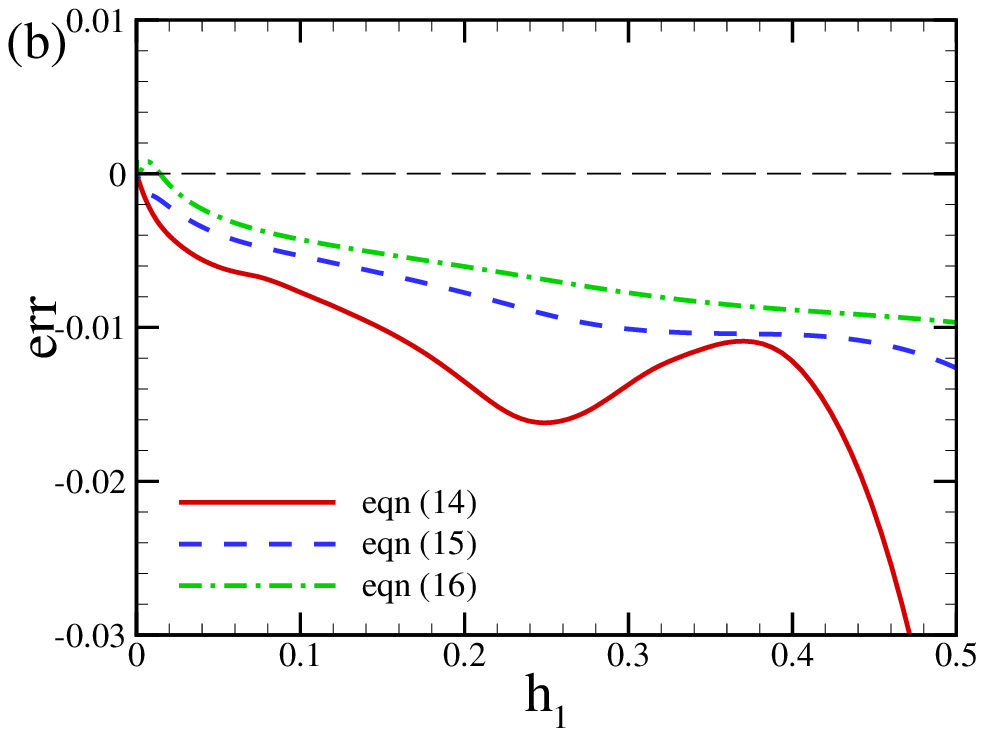}
    \caption{(a) Related errors of the skin friction coefficient estimated using equations~\eqref{eq:BL-1},~\eqref{eq:BL-2} and~\eqref{eq:BL-3}, respectively, at $Re_\theta=1410$. For equation~\eqref{eq:BL-1}, $y=h_1$. For equations~\eqref{eq:BL-2} and \eqref{eq:BL-3}, $h_0=0$; (b) the corresponding relative errors at $Re_\theta=4060$.}
    \label{fig:BL-1410-1}
\end{figure}

Figure~\ref{fig:BL-1410-1} shows the related error of the estimated skin friction coefficient using equations~\eqref{eq:BL-1},~\eqref{eq:BL-2} and~\eqref{eq:BL-3} with $h_0=0$ at $Re_\theta=1410$ (the Reynolds number based on the momentum thickness $\theta$) and 4060 respectively. It is seem that the estimated skin friction coefficient at two different locations using different methods are very close to the corresponding reference values, and the relative errors are very small. The error is relative larger for equation~\ref{eq:BL-1}, and it can be as large as to $3\%$ at $y=h_1\approx0.5$ at $Re_\theta=4060$, which demonstrates that the linear approximation of $(1-y)\tau/\tau_w$ using $f=-1.36y + 1$ will result in certain error. Nevertheless, the relative error using equations~\eqref{eq:BL-2} and~\eqref{eq:BL-3} are smaller and they are within $1\%$ at both cases.

\begin{figure}
    \centering
    \includegraphics[width=0.47\textwidth]{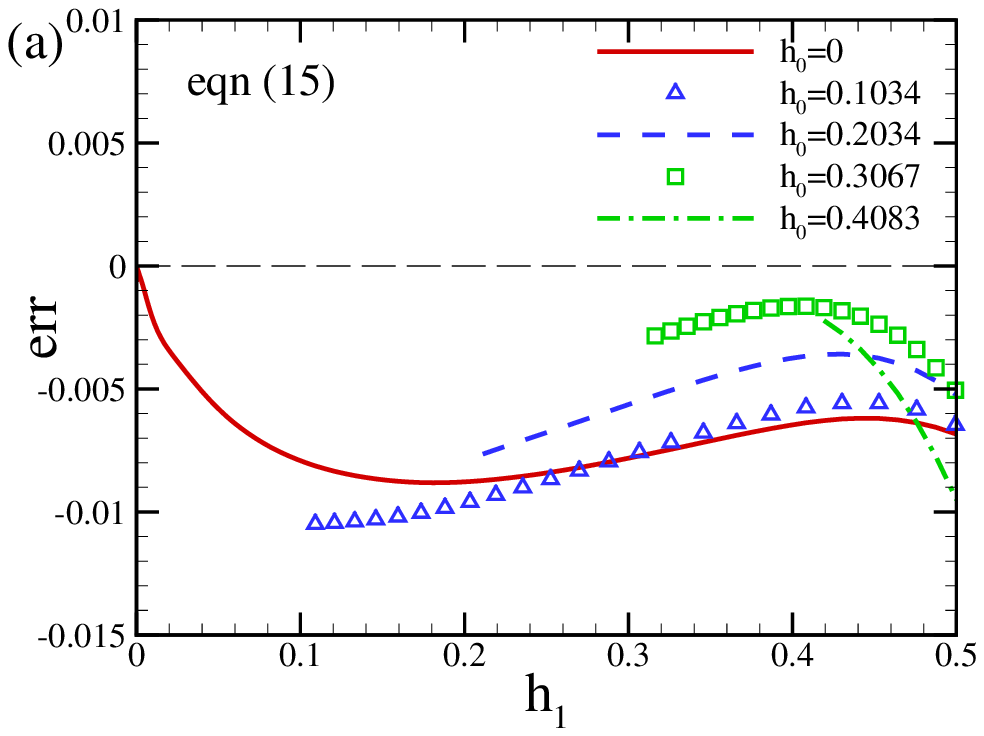}\includegraphics[width=0.47\textwidth]{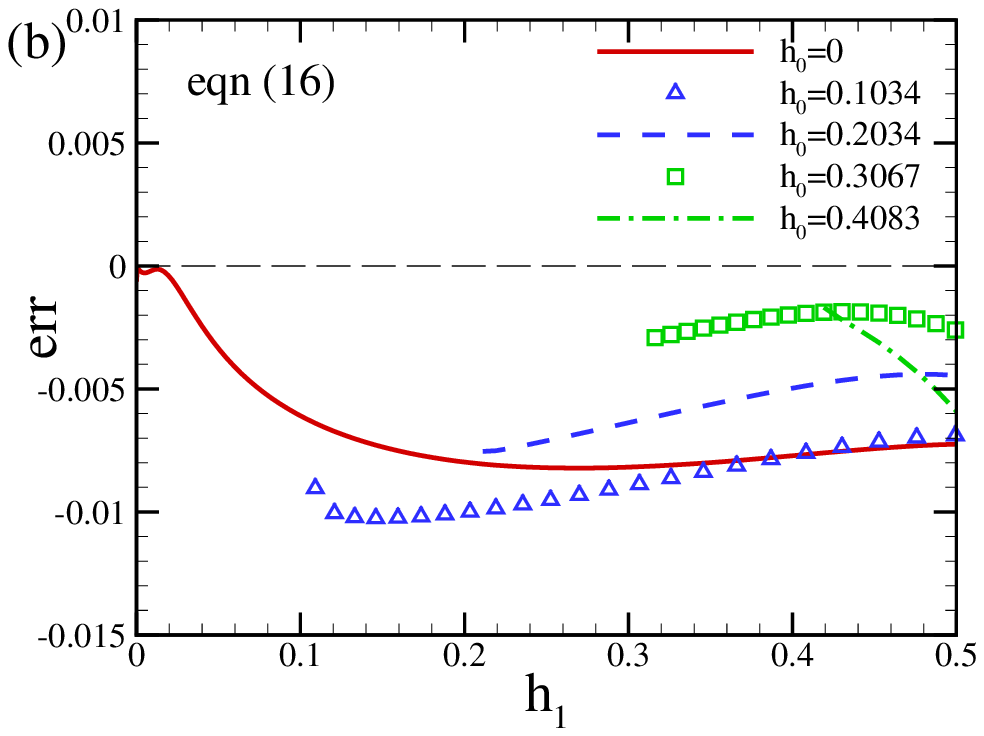}\\
    \includegraphics[width=0.47\textwidth]{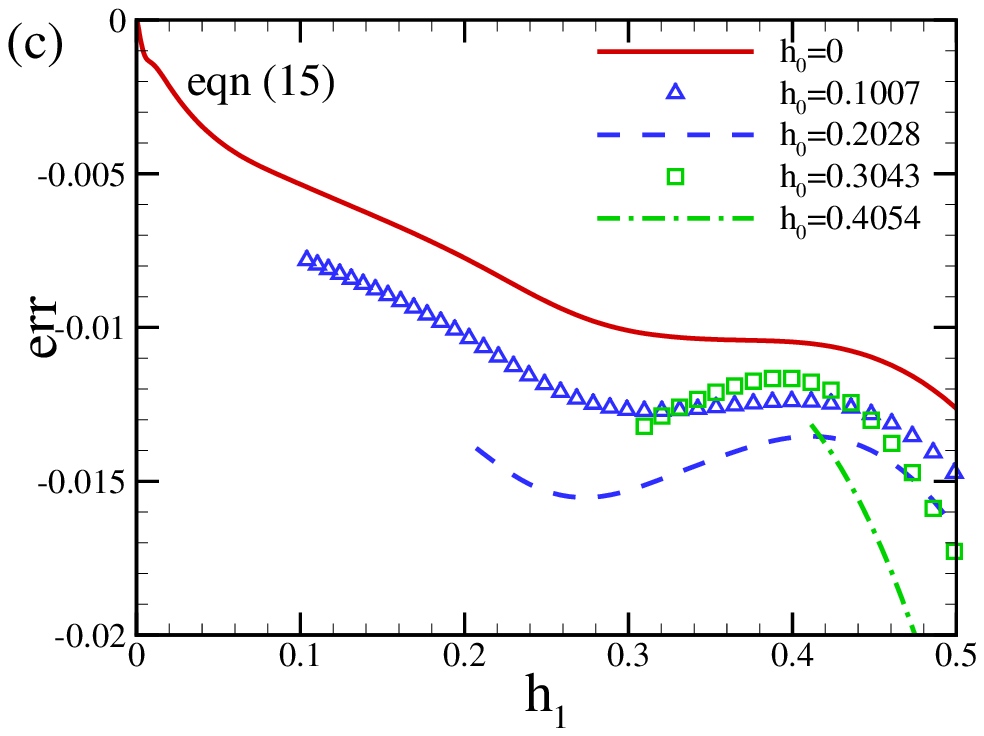}\includegraphics[width=0.47\textwidth]{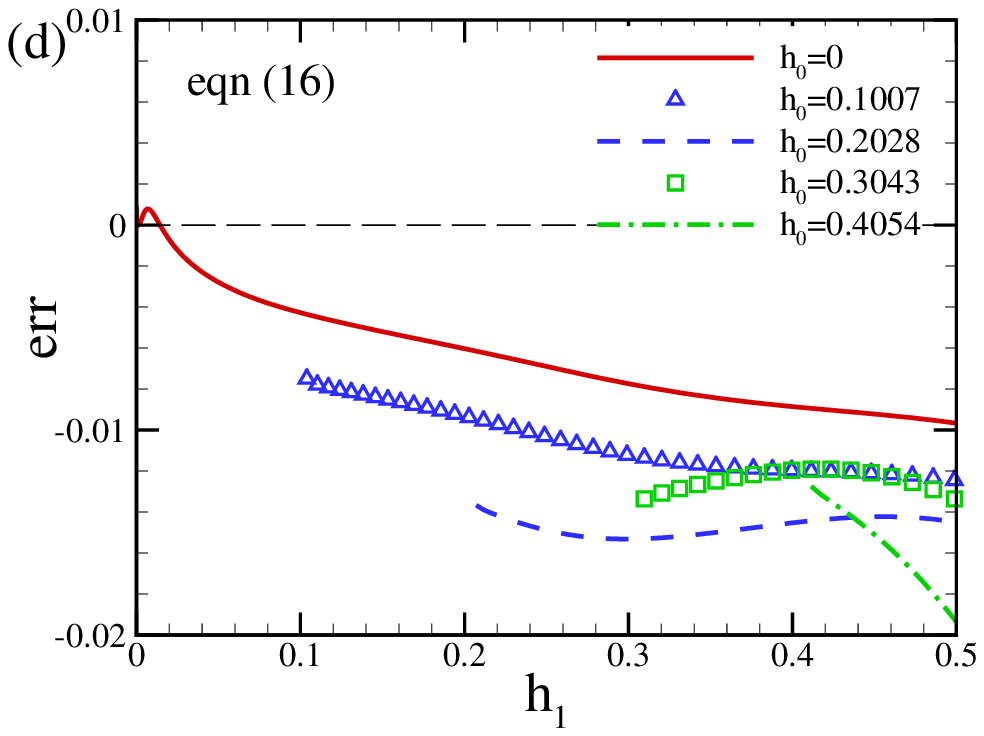}
    \caption{Related error of the skin friction coefficient estimated using equations~\eqref{eq:BL-2} and~\eqref{eq:BL-3}, respectively, with different $h_0$ at $Re_\theta=1410$ (a,b) and $Re_\theta=4060$ (c,d). (a) and (c) are using equation~\eqref{eq:BL-2}; (b) and (d) are using equation~\eqref{eq:BL-3}.}
    \label{fig:BL-1410-2}
\end{figure}

Figure~\ref{fig:BL-1410-2} shows the related error of the estimated skin friction coefficient using equations~\eqref{eq:BL-2} and~\eqref{eq:BL-3} with varying $h_0$ at $Re_\theta=1410$ and 4060 respectively. It is evident that the estimated skin friction coefficients using equations~\eqref{eq:BL-2} and~\eqref{eq:BL-3} with different $h_0$ are very close to the reference value at both locations, with relative errors less than $2\%$ for all different $h_0$. This is in sharp contrast to equation~\eqref{eq:Mehdi2} by Mehdi {\it et al.}~\cite{Mehdi2014}, where they showed that the error could be as large as $12\%$ when equation~\eqref{eq:Mehdi2} was integrated from $0.04$ to $0.5$, and that this error seems to increase if the integral starts further away from the wall. By using the empirical linear relation~\eqref{eq:BL-linear}, we could remove the requirement of the near wall information about the total stress in equation~\eqref{eq:Mehdi2} and ~\eqref{eq:XZH1}.

\begin{figure}
    \centering
    \includegraphics[width=0.48\textwidth]{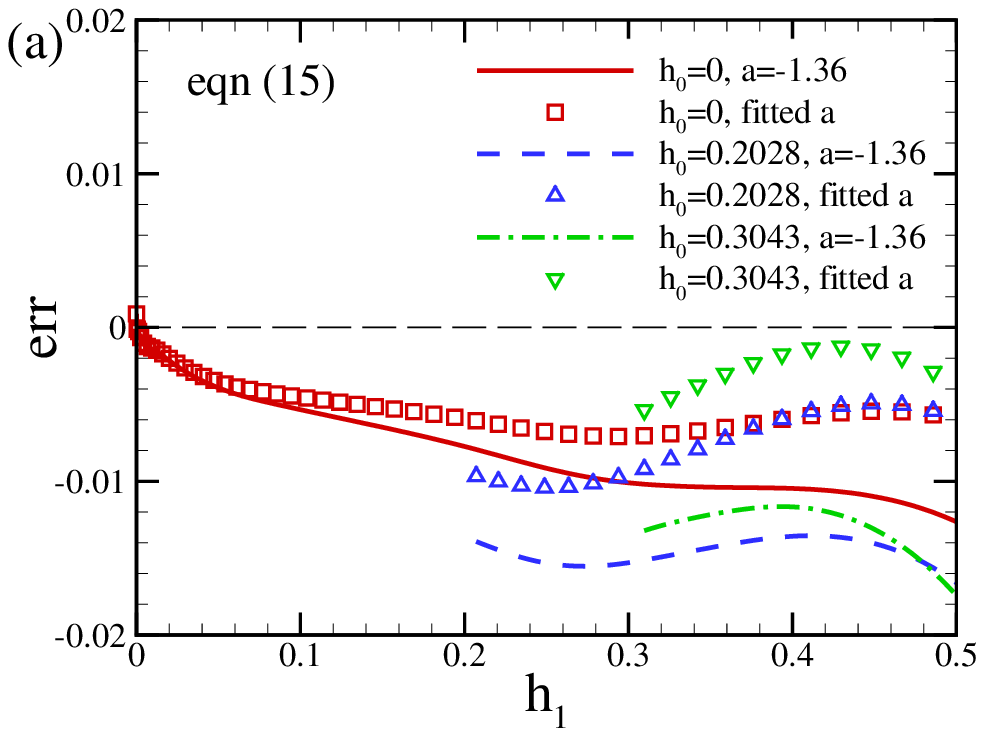}\includegraphics[width=0.48\textwidth]{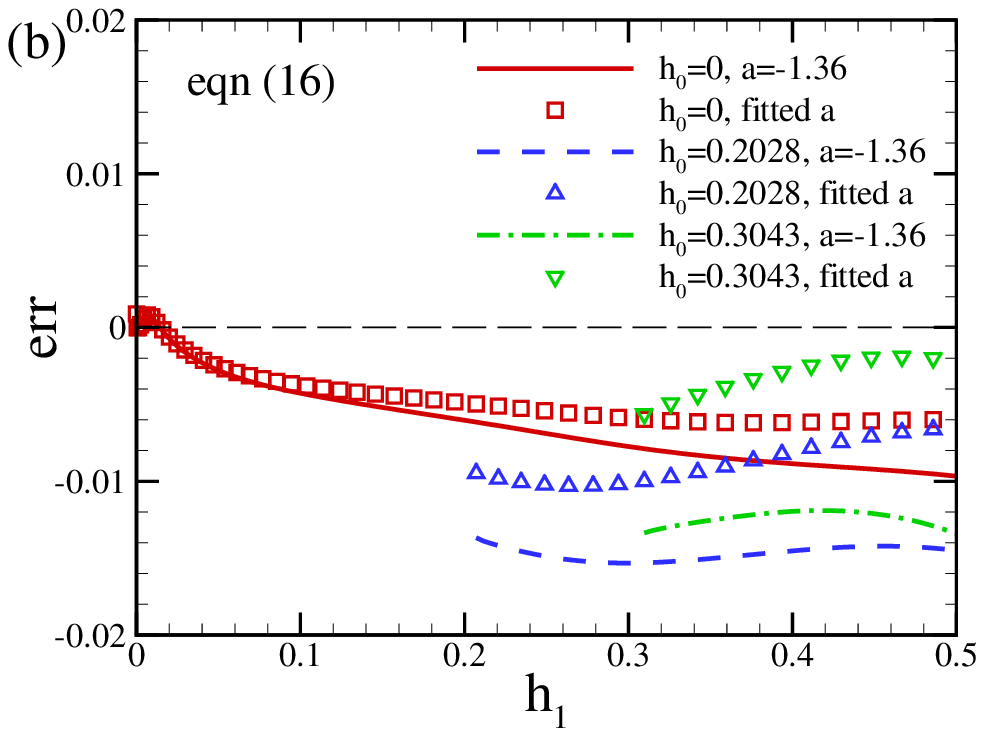}
    \caption{Related error of the skin friction coefficient estimated using (a) equation~\eqref{eq:BL-2} and (b) equation~\eqref{eq:BL-3} with different $h_0$ and $a$ at $Re_\theta=4060$. The fitted $a$ is about -1.345.}
    \label{fig:BL-4060-3}
\end{figure}

Now, we would like to investigate the influence of $a$. Figure~\ref{fig:BL-4060-3} shows the relative error of the estimated skin friction coefficient using equation~\eqref{eq:BL-2} and~\eqref{eq:BL-3} with different $h_0$ and $a$ at $Re_\theta=4060$. Here, two different values of $a$ are used, i.e. $a=-1.36$ and $a=-1.345$, where the latter is obtained by fitting the data within $y\leq0.4$. According to figure~\ref{fig:BL-4060-3}, we can see that with the more accurate fitting value of $a$, the estimated skin friction coefficients are more accurate with smaller relative errors, as expected. 

\section{Conclusions} \label{sec:Con}

In this paper, we study the skin friction in wall-bounded turbulence. Starting from the Reynolds averaged streamwise momentum equation and introducing the total stress, one could obtain two formulas for the skin friction coefficient by integrating the Reynolds averaged streamwise momentum equation from the wall to an arbitrary height twice and three times. Furthermore, if some theoretical or empirical relations for the total stress can be included, some specific formulas can be obtained without any near-wall statistics.

With the theoretical linear decay behavior of the total stress, we obtained three specific formulas to estimate the skin friction coefficient in turbulent channel flows. The formulas are validated using the direct numerical simulation data at different Reynolds numbers, and the results showed that the formulas can be quite accurate no matter where the integrating starts on the condition that the integrating region is large enough (generally larger than $0.1h$). With the empirical relation that $(1-y/\delta)\tau/\tau_w$ is linear when $y/\delta<0.5$ in turbulent boundary layer flows with zero pressure gradient, three specified formula can also be obtained. The direct numerical simulation data in turbulent boundary layer flows indeed verifies the formulas.

Because the present formulas do not require the near-wall statistics, they are well suited for the estimation of the skin friction in wall-bounded turbulence, such as turbulent channel flows and the boundary layer flows with zero pressure gradient, where the near wall statistics are very difficult to accurately obtain. The present formulas can also be used to assess the convergence of direct numerical simulation data, since the better the data converges, the smaller the relative error of the prediction is. In the future, we will extend the present derivations to the skin friction and the wall heat flux in compressible wall-bounded turbulence.

\vskip 5mm

Z.H. Xia and P. Zhang would like to thank the support by the National Natural Science Foundation of China (NSFC grant nos. 11822208, 11772297, 91852205) and the support from the Fundamental Research Funds for the central Universities.

\bibliography{SF}

\end{document}